\documentclass{article}
\usepackage{cite}
\usepackage{spconf,amsmath,amssymb,graphicx,booktabs}
\usepackage{textcase}
\usepackage{subcaption}
\usepackage[bottom]{footmisc}
\usepackage[table,xcdraw]{xcolor}
\usepackage{algorithm}
\usepackage{algpseudocode}
\usepackage{multirow}
\usepackage[disable]{todonotes}
\usepackage{tikz}
\usepackage{float}
\usepackage{paralist}
\newfloat{algorithm}{t}{lop}

\newcommand{\R}{\mathbb{R}}

\DeclareMathOperator{\Softmax}{Softmax}
\DeclareMathOperator{\Reshape}{Reshape}
\DeclareMathOperator{\DANet}{DANet}

\makeatletter
\newcommand{\algcolor}[2]{%
  \hskip-\ALG@thistlm\colorbox{#1}{\parbox{\dimexpr\linewidth-2\fboxsep}{\hskip\ALG@thistlm\relax #2}}%
}
\newcommand{\algemphY}[1]{\algcolor{yellow}{#1}}
\newcommand{\algemphR}[1]{\algcolor{pink}{#1}}
\makeatother

\setdefaultleftmargin{1.6em}{}{}{}{}{}

\newcommand{\squeezeSpace}{\vspace*{-0.25cm}}
\newcommand{\squeezeSmallSpace}{\vspace*{-0.15cm}}

\title{Closing the Training/Inference Gap for Deep Attractor Networks}
\name{Cyril Cadoux\textsuperscript{1}, Stefan Uhlich\textsuperscript{2}, Marc Ferras\textsuperscript{2} and Yuki Mitsufuji\textsuperscript{3}}
\address{\textsuperscript{1} \'{E}cole Polytechnique F\'{e}d\'{e}rale de Lausanne (EPFL), Lausanne, Switzerland\hspace*{0.54cm}\\\textsuperscript{2} Sony European Technology Center (EuTEC), Stuttgart, Germany\hspace*{2.24cm}\\\textsuperscript{3} Sony Corporation, Audio Technology Development Department, Tokyo, Japan}
\begin{document}
\ninept

\setlength{\abovedisplayshortskip}{1ex plus1ex minus1ex}
\setlength{\abovedisplayskip}{1ex plus1ex minus1ex}
\setlength{\belowdisplayshortskip}{2ex plus1ex minus1ex}
\setlength{\belowdisplayskip}{2ex plus1ex minus1ex}

\maketitle

\begin{abstract}
This paper improves the \emph{deep attractor network} (DANet) approach by closing its gap between training and inference.
During training, DANet relies on attractors, which are computed from the ground truth separations.
As this information is not available at inference time, the attractors have to be estimated, which is typically done by $k$-means.
This results in two mismatches: The first mismatch stems from using classical $k$-means with Euclidean norm, whereas masks are computed during training using the dot product similarity.
By using spherical $k$-means instead, we can show that we can already improve the performance of DANet.
Furthermore, we show that we can fully incorporate $k$-means clustering into the DANet training. This yields the benefit of having no training/inference gap and consequently results in an scale-invariant signal-to-distortion ratio (SI-SDR) improvement of 1.1dB on the Wall Street Journal corpus (WSJ0).
\end{abstract}
%
%%%%%%%%%%%%%%%%%%%%%%%%%%%%%%%%%%%%%%%%%%%%%%%%%%%%%%%%%%%%%%%%%%%%%%%%%
\section{Introduction}
\label{sec:intro}
%%%%%%%%%%%%%%%%%%%%%%%%%%%%%%%%%%%%%%%%%%%%%%%%%%%%%%%%%%%%%%%%%%%%%%%%%

Speaker-independent speech separation using \emph{deep neural networks} (DNNs) started with the pioneering work in \cite{hershey2016deep} and is currently an active research area.
It has many applications, e.g., in \emph{automatic speech recognition} (ASR) for a multi-talker scenario. Alike, such DNN approaches can be used in music separation to unmix identical instruments, e.g., first and second violins in classical recordings.

In general, there are two families of speech separation approaches that can be distinguished: approaches directly estimating the separation masks/the separated waveforms, and, approaches using an intermediate embedding from which the masks are computed.
The first family of approaches relies on the \emph{permutation-invariant training} (PIT) criterion \cite{yu2017permutation}, allowing to learn speaker-independent DNNs that work either in the \emph{short-time Fourier transform} (STFT) domain (by estimating a mask that is applied to the spectrogram \cite{yu2017permutation,kolbaek2017multitalker}) or in the time domain \cite{luo2018tasnet,luo2019conv,shi2019furcanet,shi2019end,luo2019dualpath}.
Especially the time-domain approaches have recently gained popularity as they are end-to-end, allowing to obtain better results than even oracle masks working in the STFT domain like the \emph{ideal binary mask} (IBM) or the \emph{ideal ratio mask} (IRM) \cite{luo2019conv}.
The second family of approaches projects the magnitude spectrogram into an embedding space, i.e., for each \emph{time-frequency} (T-F) bin they produce an embedding vector.
The DNNs are trained such that embeddings of T-F bins corresponding to the same speaker should be close to each other, while embeddings of different speakers should be apart.
Hence, using clustering, we can obtain a mask for each speaker, allowing to separate them. \emph{Deep clustering} (DPCL) \cite{hershey2016deep,isik2016single} was the first embedding approach, later improved by incorporating a mask inference head in the Chimera/Chimera++ network \cite{luo2017deep,wang2018alternative}.
Another embedding approach are \emph{deep attractor networks} (DANets) \cite{chen2017deep,luo2018speaker}, which are trained such that the separation error of the masks, obtained from its output embeddings, is minimized. Therefore, DANets learn the embedding implicitly.

Comparing these two families of approaches, we can note that the methods of the first family yield better separation performances as they directly optimize for the task at hand.
But they suffer from the \emph{output dimension mismatch problem} \cite{chen2017deep}, i.e., already during training we need to decide on how many speakers we want to separate.
A solution to deal with the output dimension mismatch problem for methods from the first family was recently proposed in \cite{kinoshita2018listening,takahashi2019recursive}, trying to solve the problem recursively by separating in each iteration one speaker from the remaining ones. However, they need special training whereas the methods in the second family can much more naturally deal with the output dimension mismatch problem as we only need to adapt the number of clusters $k$ at inference time in order to obtain a separation for $k$ speakers.

In this paper, we will focus on DANet, a technique working in the STFT domain that does not suffer from the output dimension mismatch problem and whose objective function directly assesses the quality of the separated sources.
We will show that DANet has the problem of a training/inference gap, which was already noted in \cite{luo2018speaker}.
Although \cite{luo2018speaker} proposes a solution, namely \emph{anchored DANet} (ADANet), the problem is not yet solved satisfactorily.
We therefore contribute two improvements for DANet in this paper: First, we propose to use a clustering based on the cosine similarity, called spherical $k$-means \cite{buchta2012spherical}, already improving the separation performance as the clustering at inference uses a similarity measure that fits to the mask computation during training.
Second, we show that we can fully incorporate the $k$-means clustering into the training, which is even better.
This has the advantage that the DANet training is aware of the clustering and, hence, we close the training/inference gap, resulting in an improved separation performance.

The paper is organized as follows: In Sec.~\ref{sec:dan}, we introduce in detail the DANet approach and discuss the problem of attractor estimation during inference. We then present two solutions to improve this estimation step in  Sec.~\ref{sec:improve_dan} and evaluate them in Sec.~\ref{sec:results}. Finally, the conclusions are drawn in Sec.~\ref{sec:conclusions}.

The following notations are used throughout this paper: $\mathbf{x}$ denotes a column vector and $\mathbf{X}$ a matrix where in particular $\mathbf{I}$ is the identity matrix. The matrix transpose, Euclidean norm, Frobenius norm and elementwise matrix product are denoted by $(.)^T$, $\lVert .\rVert$, $\lVert . \rVert_F$ and $\circ$, respectively. Finally, $\Reshape\{\mathbf{x}; N, M\}$ turns the vector $\mathbf{x}\in\R^{NM}$ into a matrix with $N$ rows and $M$ columns.

%%%%%%%%%%%%%%%%%%%%%%%%%%%%%%%%%%%%%%%%%%%%%%%%%%%%%
\section{Deep Attractor Networks (DANet)}
\label{sec:dan}
%%%%%%%%%%%%%%%%%%%%%%%%%%%%%%%%%%%%%%%%%%%%%%%%%%%%%
In this section, we will review the DANet approach \cite{chen2017deep,luo2018speaker} and highlight its training/inference gap, which we will close in Sec.~\ref{sec:improve_dan}.

Let $\mathbf{X} \in \R^{T\times F}$ denote the mixture magnitude spectrogram, consisting of $k$ speakers we want to separate and where $T$ and $F$ denote the number of time frames and frequency bins, respectively.
DANets are DNNs mapping each T-F bin of $\mathbf{X}$ to a $D$-dimensional embedding vector, i.e., $\DANet\{\mathbf{X}\} = \mathbf{V} \in \R^{TF\times D}$, where the embeddings are in the rows of $\mathbf{V}$.\\[-0.2cm]

%%%%%%%%%%%%%%%%%%%%%%%%%%%%%%%%%%%%%%%%%%%%%%%
\subsection{DANet Training}
%%%%%%%%%%%%%%%%%%%%%%%%%%%%%%%%%%%%%%%%%%%%%%%
During training, we iterate over the following steps:
\begin{compactenum}
    \item Forward propagate a mixture magnitude $\mathbf{X}$ from the training set to obtain the embeddings $\mathbf{V} = \DANet\{\mathbf{X}\}$.
    \item Compute attractors $\mathbf{a}_1,\ldots,\mathbf{a}_k$ by \begin{equation}
        \mathbf{a}_l = \frac{\mathbf{V}^T(\mathbf{u}_l\circ\mathbf{e})}{\mathbf{1}^T(\mathbf{u}_l\circ\mathbf{e})},\quad 1\le l\le k,
        \label{eq:DAN_attractor_computation}
    \end{equation}
    where $\mathbf{u}_l, \mathbf{e} \in \{0, 1\}^{TF}$ are two Boolean index vectors with $\mathbf{u}_l$ being one whenever the $l$th speaker is dominant at a specific T-F bin and zero otherwise; $\mathbf{e}$ is one for the $90\%$ most energetic T-F bins in the mixture and zero otherwise; $\mathbf{1}$ is the all-ones vector.
    In order to compute $\mathbf{u}_l$, we need the ground truth magnitude spectrograms $\mathbf{S}_1,\ldots,\mathbf{S}_k$.
    \item Obtain the masks $\mathbf{M}_l$ by
    \begin{equation}
        \mathbf{M}_l = \Reshape\{\Softmax\{\mathbf{V}\mathbf{a}_l\}; T, F\} \in \R^{T\times F},
        \label{eq:DAN_mask_computation}
    \end{equation}
    where the softmax ensures that all masks sum up to one as it is given by $[\Softmax\{\mathbf{V}\mathbf{a}_l\}]_{tf} = e^{[\mathbf{V}\mathbf{a}_l]_{tf}}/\sum_{l' = 1}^k e^{[\mathbf{V}\mathbf{a}_{l'}]_{tf}}$.
    \item Compute the \emph{mean squared error} (MSE) loss
    \begin{equation}
        L = \frac{1}{kTF}\sum_{l=1}^k \left\lVert \mathbf{S}_l - \mathbf{X} \circ \mathbf{M}_l \right\rVert_F^2
        \label{eq:DAN_formalism}
    \end{equation}
    and its gradient with respect to the DANet weights. Finally, a DNN optimizer, e.g., \emph{stochastic gradient descent} (SGD), is used to update the weights.
\end{compactenum}
These steps are repeated in a minibatch fashion and allow to learn an embedding network suited for speech separation.\\[-0.2cm]

%%%%%%%%%%%%%%%%%%%%%%%%%%%%%%%%%%%%%%%%%%%%%
\subsection{DANet Inference}
%%%%%%%%%%%%%%%%%%%%%%%%%%%%%%%%%%%%%%%%%%%%%
At inference time, we cannot compute the speaker attractors $\mathbf{a}_l$ as the Boolean index vectors $\mathbf{u}_l$ depend on the ground truth and are, hence, unknown.
We will now review the three estimation approaches proposed in \cite{chen2017deep,luo2018speaker}.\\[-0.2cm]

\noindent\emph{(E1) Fixed attractors}: Although there is no constraint on the location of the attractors, \cite{chen2017deep,luo2018speaker} empirically found that they are stable.
Therefore, one can obtain $k$ attractors for inference by using the centroids of the training attractors.\\[-0.2cm]

\noindent\emph{(E2) Attractors from $k$-means}:
We can form $k$ attractors by running $k$-means clustering \cite{duda2000pattern} on the embeddings $\mathbf{V}$ and using the found centroids $\mathbf{c}_1,\ldots,\mathbf{c}_k$ as attractors.\\[-0.2cm]

\noindent\emph{(E3) Anchored DANet (ADANet)}:
This method was proposed in \cite{luo2018speaker} as a solution to the \emph{center mismatch problem} that we will also discuss in Sec.~\ref{subsec:train_infer_gap}. Instead of assuming fixed attractors, the idea is to have $N$ trainable anchors $\mathbf{b}_1,\ldots,\mathbf{b}_N$ and to consider all its $N \choose k$ subsets of size $k$. For each subset, attractors are computed and, finally, the set of attractors with largest in-set distance is chosen.

%%%%%%%%%%%%%%%%%%%%%%%%%%%%%%%%%%%%%%%%
\subsection{Training/Inference Gap of DANet}
\label{subsec:train_infer_gap}
%%%%%%%%%%%%%%%%%%%%%%%%%%%%%%%%%%%%%%%%

We will now discuss and compare \emph{(E1)} to \emph{(E3)} for obtaining the attractors at inference time.\\[-0.2cm]

Let us first look at \emph{(E1)} and \emph{(E2)}.
Clearly, both approaches are not optimal since different approaches are taken at training (attractors are computed with \eqref{eq:DAN_attractor_computation} from the ground truth) than at inference (attractors need to be estimated).
This yields a \emph{center mismatch problem} as discussed in \cite{luo2018speaker} and results in lower separation performance than using ground truth attractors.
Furthermore, \cite{luo2018speaker} showed that \emph{(E2)} is better than \emph{(E1)}.
Comparing the training with the inference steps for \emph{(E2)} reveals that there is not only the problem that we need to estimate the attractors but also that \emph{(E2)} relies on $k$-means, which uses the Euclidean norm.
This is in contrast to the DANet training, where we computed the masks using the dot product similarity as can be seen from \eqref{eq:DAN_mask_computation}.
Therefore, \emph{(E2)} is not only suboptimal as we need to estimate the attractors but also as we are not coherent with the training.
Hence, we can improve \emph{(E2)} as we will see in Sec.~\ref{sec:spherical_kmeans_dan}.\\[-0.2cm]

ADANet \emph{(E3)} is a better approach for estimating the attractors as the same procedure for obtaining the attractors during training and inference is used.
However, compared to \emph{(E2)}, it only gives a small improvement of 0.1dB in \cite{luo2018speaker}, in our opinion due to having to choose a small number of anchors $N$ as otherwise the computational complexity is too high; another problem of ADANet is the maximum operation used to select the subset of anchors, which can be sensitive to small changes in the input mixture.
Furthermore, as already noted in \cite{luo2018speaker}, it increases considerably the computational complexity at training as well as at inference time.
In Sec.~\ref{sec:cluster_dan}, we will show a better approach directly incorporating the $k$-means into the training.
By this, we do not change the number of operations at inference time.

%%%%%%%%%%%%%%%%%%%%%%%%%%%%%%%%%%%%%%%%%%%%%%
\section{Closing the Training/Inference Gap}
\label{sec:improve_dan}
%%%%%%%%%%%%%%%%%%%%%%%%%%%%%%%%%%%%%%%%%%%%%%
In the previous section, we have seen that there is a training/inference gap for DANet, which we will now close.

%%%%%%%%%%%%%%%%%%%%%%%%%%%%%%%%%%%%%%%%%%%%%%
\subsection{DANet Inference with Spherical $\MakeTextLowercase{k}$-means}
\label{sec:spherical_kmeans_dan}
%%%%%%%%%%%%%%%%%%%%%%%%%%%%%%%%%%%%%%%%%%%%%%
\begin{figure}
    \begin{subfigure}[b]{.5\linewidth}
        \centering
        \resizebox{\linewidth}{!}{\includegraphics{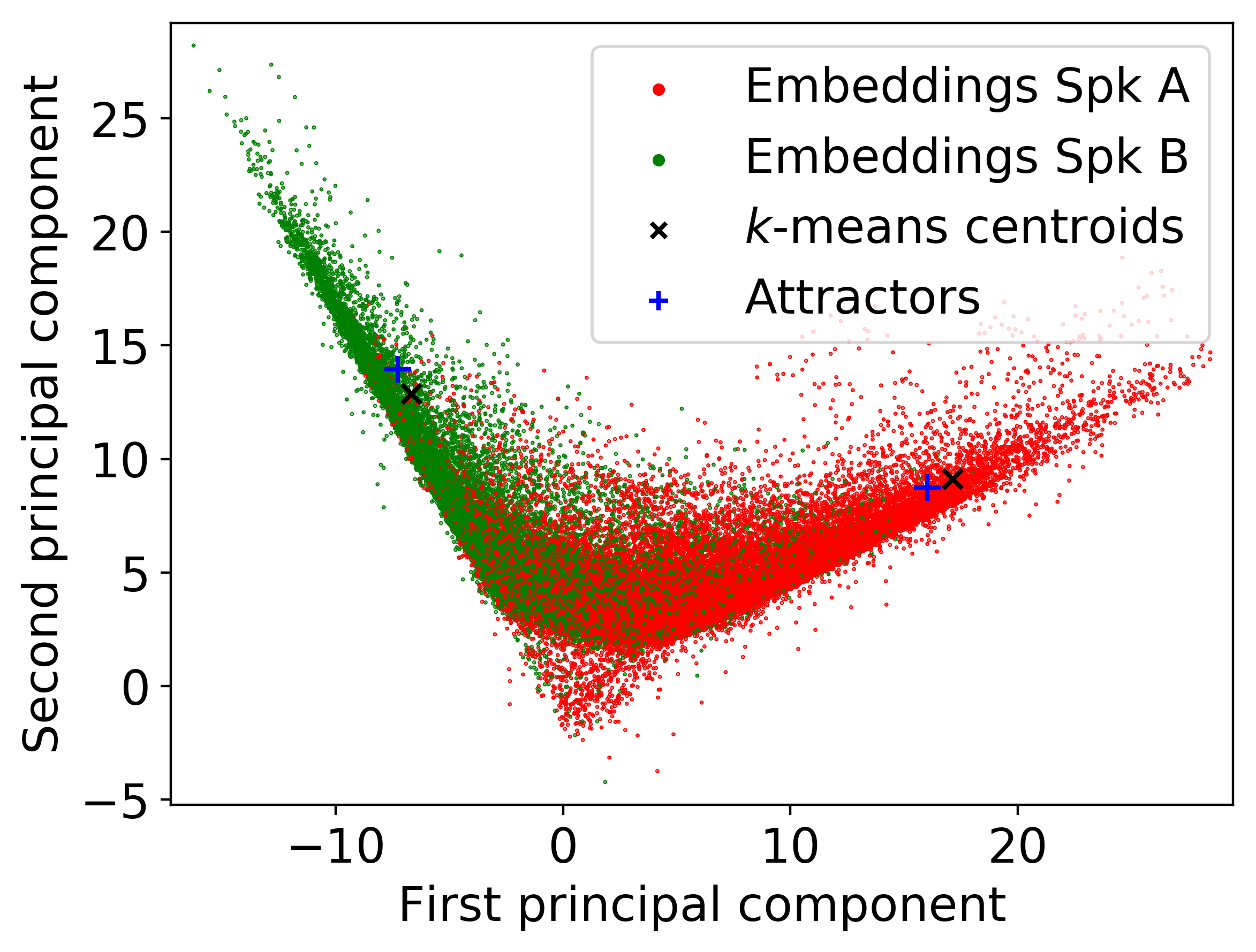}}
        \caption{Euclidean $k$-means}
        \squeezeSpace
    \end{subfigure}
    \begin{subfigure}[b]{.5\linewidth}
        \centering
        \resizebox{\linewidth}{!}{\includegraphics{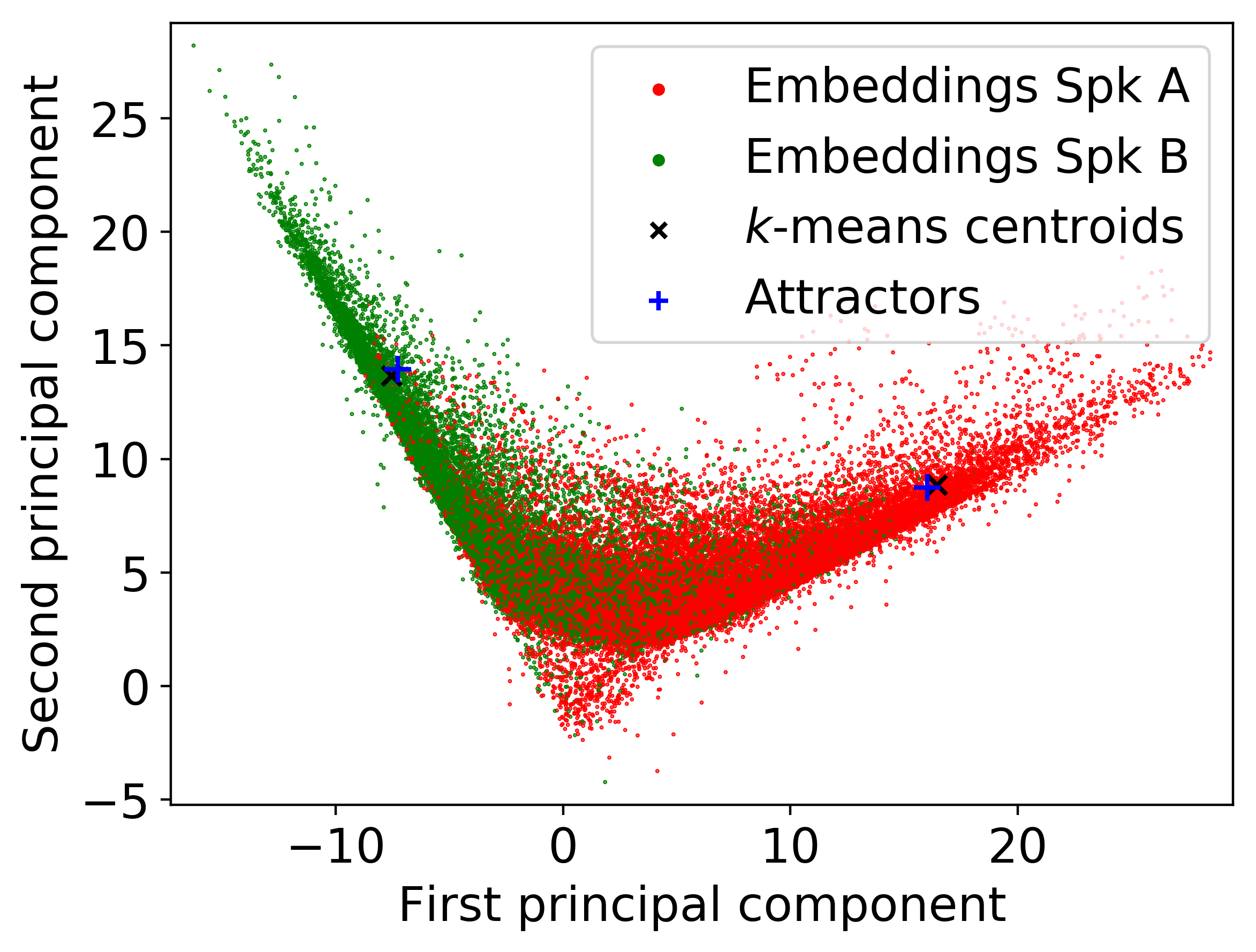}}
        \caption{Spherical $k$-means}
        \squeezeSpace
    \end{subfigure}
    \caption{PCA projection of embeddings for mix of two males. Embedding color given by TF-bin dominance of ground truth.}
    \label{fig:PCA_projections}
\end{figure}
\begin{algorithm}[t]
    \begin{algorithmic}[1]
        \State $\mathbf{\bar v}_i \leftarrow \frac{\mathbf{v}_i}{\lVert\mathbf{v}_i\rVert}$ for all $i = 1, ..., TF$
        \State Choose $k$ random centroids $\{ \mathbf{c}_1, \ldots, \ \mathbf{c}_k \}$ from $\{\mathbf{\bar v}_1, ..., \mathbf{\bar v}_{TF}\}$.
        \Repeat\par\hspace{-0.85cm}
            \algemphY{\For{$l = 1~ \mathbf{to} ~ k$}
               \Comment{Assignment step}
                \State $\mathcal{C}_l \leftarrow \{ i: \mathbf{\bar v}_i^T \mathbf{c}_l \geq \mathbf{\bar v}_i^T \mathbf{c}_m\ \ \forall \ 1 \leq m \leq k \}$
            \EndFor} \par\hspace{-0.85cm}
            \algemphR{\For{$l = 1~ \mathbf{to} ~ k$}
                \Comment{Update step}
                \State $\mathbf{c}_l \leftarrow \frac{1}{\lvert\mathcal{C}_l\rvert} \sum_{i\in\mathcal{C}_l} \mathbf{\Bar{v}}_i$
                \State $\mathbf{c}_l \leftarrow \mathbf{c}_l / \lVert\mathbf{c}_l\rVert$
            \EndFor}
        \Until{convergence}
        \State $\mathbf{c}_l \leftarrow \frac{1}{\lvert\mathcal{C}_l\rvert} \sum_{i\in\mathcal{C}_l} \mathbf{v}_i$ for all $l = 1, ..., k$
        \State \textbf{return} centroids $\{ \mathbf{c}_1, \ldots, \ \mathbf{c}_k \}$ and assignments $\{ \mathcal{C}_1, ..., \ \mathcal{C}_k \}$
    \end{algorithmic}
    \caption{Inputs: $\{\mathbf{v}_1, \mathbf{v}_2, ..., \mathbf{v}_{TF}\}$ and $k$}
    \label{alg:Spheri$k$-means}
\end{algorithm}

In order to motivate the use of spherical $k$-means, let us first study the PCA projection plot in Fig.~\ref{fig:PCA_projections}.
It shows the PCA projection of the embeddings for a mix of two male speakers.
Interestingly, we can see that the scatter diagram does not show ball-like clusters but instead clusters forming an "L" shape.
We could see this behavior for many mixtures and it indicates that the similarity measure, used to compare embeddings, should emphasize more the direction of the embeddings than their distance.

This stronger dependence on the direction is due to the mask computation in \eqref{eq:DAN_mask_computation} since it uses dot products computed by $\mathbf{V}\mathbf{a}_l$. Hence, the angle between embeddings and attractors is considered (along with its norm) as we can write $\mathbf{v}_{tf}^T\mathbf{a}_l$ as
\begin{equation}
    \mathbf{v}_{tf}^T\mathbf{a}_l = \lVert\mathbf{v}_{tf}\rVert\lVert\mathbf{a}_l\rVert\cos\angle(\mathbf{v}_{tf},\mathbf{a}_l).
\end{equation}
Classical $k$-means as used in \emph{(E2)} is not a good fit for such an embedding distribution since it relies on the Euclidean norm \cite{duda2000pattern}. Hence, using classical $k$-means will yield attractors that are further away from the ideal ones computed in \eqref{eq:DAN_attractor_computation}.

Therefore, we propose to use spherical $k$-means\cite{buchta2012spherical} at inference time to approximate attractors in a way that remains coherent with the DANet training.
The algorithm is outlined in Alg.~\ref{alg:Spheri$k$-means} and it differs from Euclidean $k$-means in the assignment step which uses the cosine similarity instead of minimizing Euclidean distance between observations and centroids. In Fig.~\ref{fig:PCA_projections}, we can see that the spherical $k$-means centroids are closer to the ideal attractors than the ones from classical $k$-means. 

In Sec.~\ref{sec:results} we will compare the performance of the same network evaluated with spherical $k$-means clustering and Euclidean $k$-means clustering, showing that spherical $k$-means gives a $0.3$dB improvement in SI-SDR. 
Please note that this improvement can be used for any already trained DANet as we only need to change the inference clustering from $k$-means to spherical $k$-means.

%%%%%%%%%%%%%%%%%%%%%%%%%%%%%%%%%%%%%%%%%%%%%%%%%%%%%%%%%%%%%%%%%%%%%%%%%
\subsection{$\MakeTextLowercase{k}$-means DANet}
\label{sec:cluster_dan}
%%%%%%%%%%%%%%%%%%%%%%%%%%%%%%%%%%%%%%%%%%%%%%%%%%%%%%%%%%%%%%%%%%%%%%%%%
\begin{figure}
    \centering
    \includegraphics[width=\linewidth]{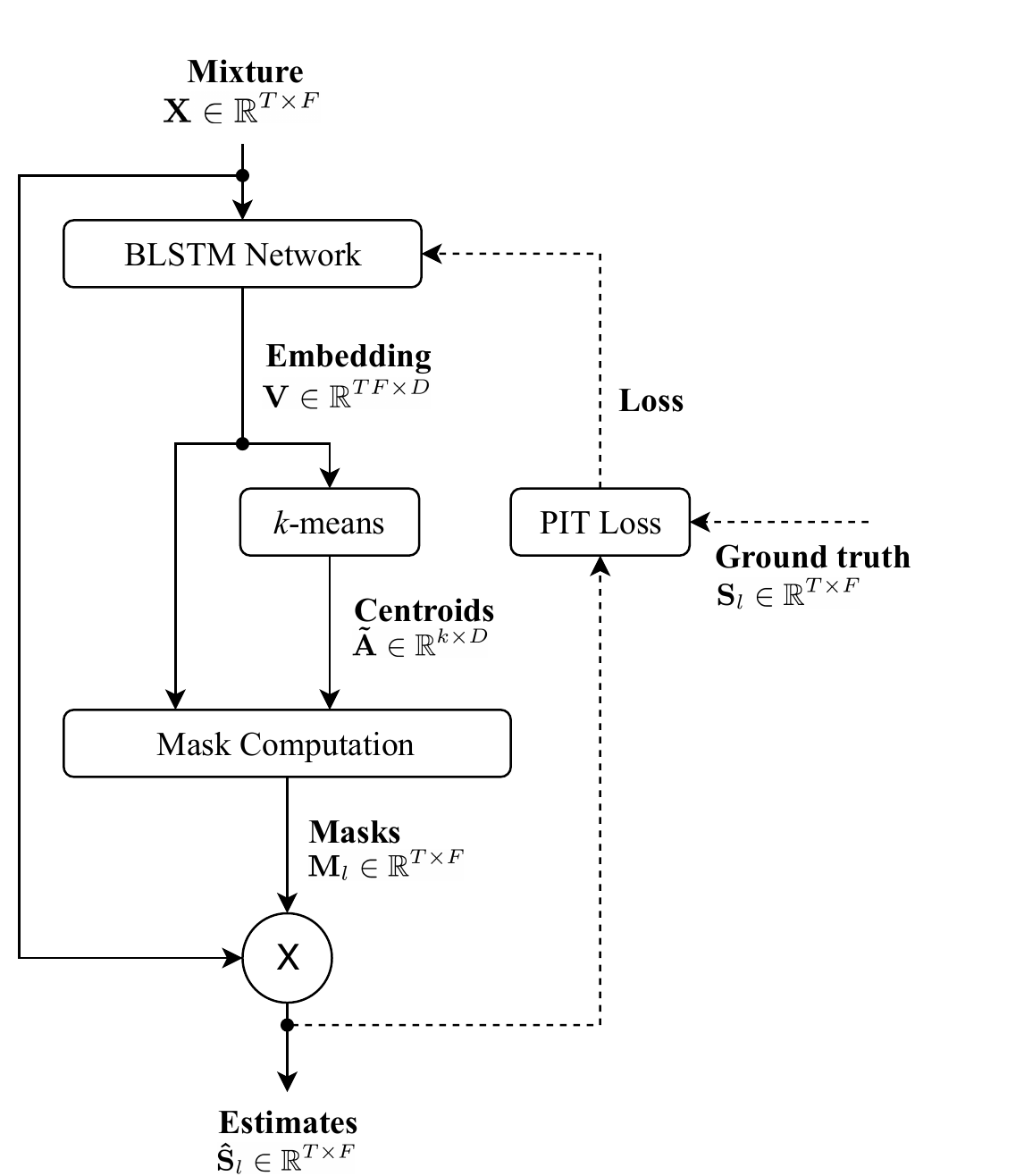}
    \caption{Architecture of $k$-means DANet.}
    \label{fig:architecture}
\end{figure}

Despite solving the metric mismatch and, hence, improving speech separation results, using spherical $k$-means does not help to solve the deeper training/inference gap, representing the main limitation of DANet.

To keep training consistent with inference, we should use the same procedure to obtain the attractors during both phases.
Although ADANet tried to achieve this, the problem is not yet solved satisfactorily as discussed in Sec.~\ref{subsec:train_infer_gap}.
Hence, we propose to introduce $k$-means into the training by replacing attractors with $k$-means centroids.
The overall architecture of our framework is shown in Fig.~\ref{fig:architecture} and we will refer to it in the following as \emph{$k$-means DANet}.

In order to add $k$-means to the network, we unfold $k$-means into $L$ iterations such that they are now part of the computational graph through which we perform the forward and backward passes to train the network \cite{hershey2014deep}. As the `argmin`/`argmax` operations in the assignment step are non-differentiable, gradients will flow only through the centroids computed in the update step (cf. Alg.~\ref{alg:Spheri$k$-means}). This means, we assume that the assignment of embeddings to a cluster will not change for small changes in the network weights, which is a reasonable assumption.

Note that we can now use either the classical or spherical version of $k$-means. In order to adjust the mask computation to the version of $k$-means, we use the following formulas to compute the masks:\\[-0.2cm]

\noindent\emph{Classical $k$-means}:
\begin{subequations}
\begin{equation}
    \label{eq:Euclid_$k$-means_DAN_mask_computation}
    [\mathbf{M}_l]_{tf} = \Softmax(-\lVert \mathbf{v}_{tf} - \mathbf{c}_l \rVert).
\end{equation}

\noindent\emph{Spherical $k$-means}:
\begin{equation}
    \label{eq:Shperi_$k$-means_DAN_mask_computation}
    [\mathbf{M}_l]_{tf} = \Softmax(\mathbf{v}_{tf}^T \mathbf{c}_l).
\end{equation}
\end{subequations}
As the masks are only derived from the result of the $k$-means clustering, the mapping between ground truth and estimated utterances is unknown. Therefore, we need to use the PIT \cite{yu2017permutation} criterion for training, which computes all $k!$ possibilities and selects the one with the smallest MSE.

It is interesting to compare $k$-means DANet to DPCL++ \cite["end-to-end"]{isik2016single}, an improved version of DPCL, augmenting it with unfolded soft clustering layers and a further mask enhancement network. Although this approach makes the training aware of the clustering, it does not ensure that the embeddings after the fine-tuning of the full network stack are still good embedding vectors (e.g., allow to use clustering to infer a good separation mask) due to the enhancement network. This is in contrast to our approach which ensures that the embeddings are well suited for mask inference.

%In summary, as the clustering is now part of the training, we will learn embeddings that are well-suited for clustering.

\begin{table}
    \centering
    %\resizebox{\linewidth}{!}{
    \scalebox{0.95}{
    \begin{tabular}{lcc }
        \toprule
        \textbf{Network} & \textbf{Inference $k$-means} &
        \textbf{SI-SDRi} \\
        \midrule
         DANet (our implementation) & Euclidean & $9.6$ \\ % runx4_2019Apr10-h14m55s18_2speakers8k_DAN_DOT_BASIC_realTR_newtry_MEAN_L2eval_keep90/2speakers/wav8k/min/tt/ : 10.5098 (10.2602) // 9.6393 (9.6421) dB
         DANet (our implementation) & Spherical & $9.9$ \\ % [KEEP 90% TRAINING][THRESH KMEANS] runx4_2019Apr10-h14m55s18_2speakers8k_DAN_DOT_BASIC_realTR_newtry_MEAN_SHPERIeval_keep90/2speakers/wav8k/min/tt/ : 10.6872 (10.4376) // 9.9071 (9.9099) dB
         % The two different evaluations come from a model named runx4_2019Apr10-h14m55s18_2speakers8k_DAN_DOT_BASIC_realTR_newtry I think. The name in the SDR_results file are extended to show which evaluation was used
         %\midrule
         %DANet (w/o CL, from \cite[Table 1]{luo2018speaker}) & Euclidean & $9.5$ \\ % Table I, row DANet-90% Softmax
         %DANet (w/$\phantom{o}$ CL, from \cite[Table 1]{luo2018speaker}) & Euclidean & $10.3$ \\ % Table I, row DANet-90% Softmax
        \bottomrule
    \end{tabular}
    }
    \squeezeSmallSpace
    \caption{SI-SDR improvements (dB) on wsj0-2mix using DANet where inference is done with different $k$-means metrics.}
    \label{tab:DAN_improvement_spheriEval}

\vspace{0.3cm}

    \centering
    \resizebox{\linewidth}{!}{
    \begin{tabular}{lccccc}
        \toprule
        \textbf{$k$-means unfold} & $L = 1$ & $L = 3$ & $L = 5$ & $L = 10$ & $L = 20$ \\
        \midrule
        \textbf{SI-SDRi} & 9.7 & 10.2 & 10.5 & 10.7 & 10.7\\
        \textbf{Training time} & \multirow{2}{*}{4.60} & \multirow{2}{*}{6.51} & \multirow{2}{*}{8.54} & \multirow{2}{*}{13.45} & \multirow{2}{*}{22.81}\\
        (one epoch, in minutes)\\
        \bottomrule
    \end{tabular}}
    \squeezeSmallSpace
    \caption{SI-SDRi (dB) on wsj0-2mix for $k$-means DANet for different number $L$ of unfolded $k$-means iterations used during training.}
    \label{tab:unfold_iter}
\end{table}

% put table here as two column tables are always put on the next page
\begin{table*}
    \centering
    \resizebox{0.95\linewidth}{!}{
    \begin{tabular}{lccccccccc}
        \toprule
        \multirow{2}{*}{\textbf{Method}}
        & \hspace{0.2cm}
        & \multicolumn{2}{c}{\textbf{Trained on wsj0-2mix}}
        & \hspace{0.2cm}
        & \multicolumn{2}{c}{\textbf{Trained on wsj0-3mix}}
        & \hspace{0.2cm}
        & \multicolumn{2}{c}{\textbf{Trained on wsj0-23mix}} \\
        &
        & $k = 2$ Spkr & $k = 3$ Spkr
        &
        & $k = 2$ Spkr & $k = 3$ Spkr
        &
        & $k = 2$ Spkr & $k = 3$ Spkr\\
        \midrule
         DANet \cite{luo2018speaker} (w/o CL) & & 9.5  & - & & - & - & & - & - \\ % Table I row DANet-90% Sigmoid
         
         DANet \cite{luo2018speaker} (w/$\phantom{o}$ CL) & & 10.0 & - & & - & 8.6 & & - & - \\ % Table I row DANet-90%* // Table IV row DANet-Kmeans* col SI-SNRi // Table V row DANet-Kmeans* col 2 Spk // col 3 Spk
         
         \midrule
         ADANet \cite{luo2018speaker} (w/o CL) & & 9.6 & - & & - & - & & - & - \\ % Table II row "6"
         
         ADANet \cite{luo2018speaker} (w/$\phantom{o}$ CL) & & 10.4 & - & & - & \textbf{9.1} & & - & - \\ % Table II row 6-do* // Table IV row ADANet-6-do* col SI-SNRi // Table V row ADANet-6-do* col 2 Spk // col 3 Spk
         
         \midrule
         DPCL \cite{isik2016single} & & 10.3 & 2.1 & & 8.5 & 7.1 & &\textbf{ 10.5} & 7.1 \\ % Table 7 row end-to-end col overall // Table 5 row 2 speaker col 3speaker // row 3 speaker col 2 speaker // row 3 speaker col 2 speaker // row Mixed curriculum col 2 speaker // row Mixed curriculum col 3 speaker
         DPCL++ \cite{isik2016single} & & \textbf{10.8} & - & & - & - & & - & -\\
         \midrule 
         $k$-means DANet$\phantom{^+}$ (Euclidean)& & 10.5 & 3.0 & & 7.7 & 7.6 & & 10.2 & \textbf{8.3} \\
         % runx4_2019May03-h16m29s22_2speakers8k_DAN_L2_KMEANS_energyProp_realTR/2speakers/wav8k/min/tt/ 
         % runx4_2019May03-h16m29s22_2speakers8k_DAN_L2_KMEANS_energyProp_realTR/3speakers/wav8k/min/tt/ 
         % runx4_2019Jun14-h13m18s00_3speakers8k_DAN_L2_KMEANS_energyPropz_MEAN/2speakers/wav8k/min/tt/ 
         % runx4_2019Jun14-h13m18s00_3speakers8k_DAN_L2_KMEANS_energyPropz_MEAN/3speakers/wav8k/min/tt/ 
         % runx4_2019Aug07-h10m16s11_23speakers8k_DAN_L2_KMEANS_MEAN/2speakers/wav8k/min/tt/ 
         % runx4_2019Aug07-h10m16s11_23speakers8k_DAN_L2_KMEANS_MEAN/3speakers/wav8k/min/tt/ 
         
         $k$-means DANet$^+$ (Euclidean) & & \textbf{10.8} & 3.1 & & \textbf{8.9} & 8.4 & & 10.4 & 8.2 \\
         % runx4_2019Jun14-h10m40s59_DAN_L2_KMEANS_energyProp_10iter_long_MEAN/2speakers/wav8k/min/tt/ 
         % runx4_2019Jun14-h10m40s59_2speakers8k_DAN_L2_KMEANS_energyProp_10iter_long_MEAN/3speakers/wav8k/min/tt/ 
         % runx4_2019Aug07-h14m25s06_STEFAN_3speakers8k_DAN_L2_KMEANS_LONG_MEAN/2speakers/wav8k/min/tt/ 
         % runx4_2019Aug07-h14m25s06_STEFAN_3speakers8k_DAN_L2_KMEANS_LONG_MEAN/3speakers/wav8k/min/tt/ 
         % runx4_2019Aug12-h08m55s28_STEFAN_23speakers8k_DAN_L2_KMEANS_LONG_MEAN/2speakers/wav8k/min/tt/ 
         % runx4_2019Aug12-h08m55s28_STEFAN_23speakers8k_DAN_L2_KMEANS_LONG_MEAN/3speakers/wav8k/min/tt/ 
         
         \midrule 
         $k$-means DANet$\phantom{^+}$ (spherical) & & 10.5 & \textbf{3.6} & & 7.7 & 7.8 & & 10.1 & 7.8 \\
         % 2speakers8k_DAN_DOT_KMEANS_energyProporCONTINUED_MEAN/2speakers/wav8k/min/tt/
         % 2speakers8k_DAN_DOT_KMEANS_energyProporCONTINUED_MEAN/3speakers/wav8k/min/tt/ 
         % runx4_2019Jul31-h13m05s28_3speakers8k_DAN_DOT_KMEANS_simple_rdminit_MEAN/2speakers/wav8k/min/tt/ 
         % runx4_2019Jul31-h13m05s28_3speakers8k_DAN_DOT_KMEANS_simple_rdminit_MEAN/3speakers/wav8k/min/tt/ 
         % runx4_2019Aug06-h15m57s13_23speakers8k_DAN_DOT_KMEANS_MEAN/2speakers/wav8k/min/tt/
         % runx4_2019Aug06-h15m57s13_23speakers8k_DAN_DOT_KMEANS_MEAN/3speakers/wav8k/min/tt/ 
         
         $k$-means DANet$^+$ (spherical) & & \textbf{10.8} & 3.2 & & 7.6 & 7.6 & & 10.3 & 8.0 \\
         % runx4_2019Jun24-h16m27s03_2speakers8k_DAN_DOT_KMEANS_energyProp_10iter_longg_MEAN/2speakers/wav8k/min/tt/ 
         % runx4_2019Jun24-h16m27s03_2speakers8k_DAN_DOT_KMEANS_energyProp_10iter_longg_MEAN/3speakers/wav8k/min/tt/ 
         % runx4_2019Aug07-h14m26s19_STEFAN_3speakers8k_DAN_DOT_KMEANS_LONG_MEAN/2speakers/wav8k/min/tt/ 
         % runx4_2019Aug07-h14m26s19_STEFAN_3speakers8k_DAN_DOT_KMEANS_LONG_MEAN/3speakers/wav8k/min/tt/ 
         % runx4_2019Aug12-h08m46s44_STEFAN_23speakers8k_DAN_DOT_KMEANS_LONG_MEAN/2speakers/wav8k/min/tt/ 
         % runx4_2019Aug12-h08m46s44_STEFAN_23speakers8k_DAN_DOT_KMEANS_LONG_MEAN/3speakers/wav8k/min/tt/ 
         
        \bottomrule
    \end{tabular}
    }
    \caption{SI-SDR improvements (dB) for two and three speakers mixtures for $k$-means DANet trained on different datasets.}
    \label{tab:DAN_perf}
\end{table*}

%%%%%%%%%%%%%%%%%%%%%%%%%%%%%%%%%%%%%%%%%%%%%%%%%
\section{Results}
\label{sec:results}
%%%%%%%%%%%%%%%%%%%%%%%%%%%%%%%%%%%%%%%%%%%%%%%%%
We will now evaluate our proposed improvements from Sec.~\ref{sec:improve_dan}.

%%%%%%%%%%%%%%%%%%%%%%%%%%%%%%%%%%%%%%%%%%%%%%%%%
\subsection{Setup}
\label{subsec:results_setup}
%%%%%%%%%%%%%%%%%%%%%%%%%%%%%%%%%%%%%%%%%%%%%%%%%
We use the corpus introduced in \cite{hershey2016deep} which is built from the \emph{Wall Street Journal} (WSJ0) dataset.
The wsj0-2mix dataset is constructed for two-speaker separation by selecting pairs of utterances from different speakers in WSJ0 and mixing them with a random SNR between $-5$ dB and $5$ dB.
In the end, wsj0-2mix consists of a 30h training set, a 10h validation set, and a 5h testing set.
The wsj0-3mix dataset is constructed similarly by mixing utterances from 3 different speakers.
Finally, we also construct the wsj0-23mix dataset by merging wsj0-2mix and wsj0-3mix into a single dataset.

Input data are formed by resampling every mixture to $8$kHz and taking the magnitude of their STFT computed with Hanning windows of $64$ms and $75$\% overlap. Embeddings have a dimension of $D = 20$.

We use the same network architecture as described in \cite{stoter19}, an open-source implementation of \cite{uhlich2017improving}, where we only modify the output shape to obtain embeddings of size $D$ for each T-F bin.
Three \emph{bi-directional LSTM} (BLSTM) layers are preceded by two fully-connected layers that encode the input frames and are succeeded by two fully-connected layers that decode to the desired output dimension. Each BLSTM layer has 256 cells.
Unless stated otherwise, our networks are trained for $350$ epochs with Adam \cite{kingma2014adam} and an initial learning rate of $10^{-3}$, which is divided by 3 (resp. 10, 30, 100) after 150 (resp. 225, 300, 325) epochs. Please note that we use the same setup (network architecture, training settings) for all our experiments except for the DANet$^+$ in Table~\ref{tab:DAN_perf}. This allows us to do fair comparisons and to see the difference due to our improvements from Sec.~\ref{sec:improve_dan}.

Furthermore, we modified $k$-means to take into account T-F bin energies as proposed in \cite{isik2016single}. More specifically, we compute clusters~as
\begin{equation}
    \textstyle\mathbf{c}_l = (\sum\nolimits_{i \in \mathcal{C}_l} [\mathbf{X}]_i^2\mathbf{v}_i) / (\sum\nolimits_{i \in \mathcal{C}_l} [\mathbf{X}]_i^2).
\end{equation}
This ensures that centroids are closer to embeddings with high-energy, i.e., closer to important bins with a higher influence on the separation error. We use this weighted version of $k$-means for all experiments in this paper.

We evaluate our models using the \emph{scale-invariant signal-to-distortion ratio improvement} (SI-SDRi), as described in \cite{SDR_paper,le2019sdr}.

%%%%%%%%%%%%%%%%%%%%%%%%%%%%%%%%%%%%%%%%%%%%%%%%%%%%%%%%
\subsection{Results for DANet Inference with Spherical $k$-means}
\label{subsec:results_spheri_DAN_evaluation}
%%%%%%%%%%%%%%%%%%%%%%%%%%%%%%%%%%%%%%%%%%%%%%%%%%%%%%%%
Table~\ref{tab:DAN_improvement_spheriEval} gives the speech separation performance of DANet if we use classical $k$-means or the proposed spherical $k$-means. We can see that spherical $k$-means can increase SI-SDRi from $9.6$dB to $9.9$dB, i.e., we can improve by $0.3$dB. Please note that we apply the same DANet and only modified the clustering algorithm used during inference, i.e., the $0.3$dB improvements are solemnly due to using spherical $k$-means.

%For comparison, we also give in Table~\ref{tab:DAN_improvement_spheriEval} the best results from \cite{luo2018speaker}, which were obtained by combining \emph{curriculum learning} (CL) with Dropout. We can expect that using spherical $k$-means would also improve the separation performance of these networks.

%%%%%%%%%%%%%%%%%%%%%%%%%%%%%%%%%%%%%%%%%%%%%%%%
\subsection{Results for $k$-means DANet}
\label{subsec:results_$k$-means_DAN}
%%%%%%%%%%%%%%%%%%%%%%%%%%%%%%%%%%%%%%%%%%%%%%%%
We first analyze the influence of the $k$-means unfolding iterations $L$ on the performance. Table~\ref{tab:unfold_iter} gives the results and we can see that increasing $L$ is beneficial as expected but that the improvement saturates for larger $L$ values. In particular, a value of $L = 10$ is sufficient. For completeness, Table~\ref{tab:unfold_iter} also contains the training times with a RTX 2080 Ti for one epoch in minutes and we can see that increasing $L$ leads to moderate increases of the training time.

Comparing the results of Table~\ref{tab:DAN_improvement_spheriEval} with Table~\ref{tab:unfold_iter}, we can see that $k$-means DANet performs considerably better than DANet. For example, with $L = 10$ we have 10.7dB SI-SDRi which is 1.1dB better than DANet with classical $k$-means and 0.8dB better than a DANet with spherical $k$-means used during inference. This is a considerable improvement, showing that closing the gap between training and inference is important. Please note again that the only difference is the addition of the unfolded $k$-means iterations to the network; in particular, we did not modify the number of trainable weights and the improvements that we can see is solemnly due to making the training of DANet aware of the $k$-means clustering.

Finally, Table~\ref{tab:DAN_perf} gives speech separation results for networks trained with wsj0-2mix, wsj0-3mix and wsj0-23mix.
Unless stated otherwise, it is assumed for our $k$-means DANet models that $L=5$ $k$-means iterations are used during training and that $20$ iterations are used at inference time.
Notation "$^+$" denotes models trained for $700$ epochs (instead of $350$) and with $L = 10$ unfolded $k$-means iterations.
We compare our networks to state-of-the-art embedding based models that do not rely on phase processing: DPCL~\cite{isik2016single}, DPCL++~\cite{isik2016single}, which combines an enhancement network with soft-clustering, DANet~\cite{luo2018speaker} and ADANet~\cite{luo2018speaker}. For DANet/ADANet, we give the results for training with/without \emph{curriculum learning} (CL) \cite{luo2018speaker}.
The results in Table~\ref{tab:DAN_perf} show that $k$-means DANet achieves state-of-the-art SI-SDRi results.

%%%%%%%%%%%%%%%%%%%%%%%%%%%%%%%%%%%%%%%%%%%%%%
\section{Conclusions}
\label{sec:conclusions}
%%%%%%%%%%%%%%%%%%%%%%%%%%%%%%%%%%%%%%%%%%%%%%

In this work, we proposed two modifications of \emph{deep attractor networks} (DANet) aiming at removing the gap between training and evaluation.
We could show that both improve the performance.
In particular, replacing attractors derived from ground truth by $k$-means centroids during training is an efficient way to improve the DANet performance as it completely removes any gap between training and inference.

\bibliographystyle{IEEEbib}
\bibliography{kmeans}

\begin{thebibliography}{10}

\bibitem{hershey2016deep}
J.~R. Hershey, Z.~Chen, J.~Le~Roux, and S.~Watanabe,
\newblock ``Deep clustering: Discriminative embeddings for segmentation and
  separation,''
\newblock in {\em International Conference on Acoustics, Speech and Signal
  Processing (ICASSP)}, 2016, pp. 31--35.

\bibitem{yu2017permutation}
D.~Yu, M.~Kolb{\ae}k, Z.-H. Tan, and J.~Jensen,
\newblock ``Permutation invariant training of deep models for
  speaker-independent multi-talker speech separation,''
\newblock in {\em International Conference on Acoustics, Speech and Signal
  Processing (ICASSP)}, 2017, pp. 241--245.

\bibitem{kolbaek2017multitalker}
M.~Kolb{\ae}k, D.~Yu, Z.-H. Tan, and J.~Jensen,
\newblock ``Multitalker speech separation with utterance-level permutation
  invariant training of deep recurrent neural networks,''
\newblock {\em IEEE/ACM Transactions on Audio, Speech and Language Processing
  (TASLP)}, vol. 25, no. 10, pp. 1901--1913, 2017.

\bibitem{luo2018tasnet}
Y.~Luo and N.~Mesgarani,
\newblock ``Tasnet: time-domain audio separation network for real-time,
  single-channel speech separation,''
\newblock in {\em 2018 IEEE International Conference on Acoustics, Speech and
  Signal Processing (ICASSP)}, 2018, pp. 696--700.

\bibitem{luo2019conv}
Y.~Luo and N.~Mesgarani,
\newblock ``{Conv-TasNet}: Surpassing ideal time--frequency magnitude masking
  for speech separation,''
\newblock {\em IEEE/ACM Transactions on Audio, Speech, and Language
  Processing}, vol. 27, no. 8, pp. 1256--1266, 2019.

\bibitem{shi2019furcanet}
Z.~Shi, H.~Lin, L.~Liu, R.~Liu, and J.~Han,
\newblock ``{FurcaNet}: An end-to-end deep gated convolutional, long short-term
  memory, deep neural networks for single channel speech separation,''
\newblock {\em arXiv preprint arXiv:1902.00651}, 2019.

\bibitem{shi2019end}
Z.~Shi, H.~Lin, L.~Liu, R.~Liu, S.~Hayakawa, S.~Harada, and J.~Han,
\newblock ``End-to-end monaural speech separation with multi-scale dynamic
  weighted gated dilated convolutional pyramid network,''
\newblock {\em Proc. Interspeech 2019}, pp. 4614--4618, 2019.

\bibitem{luo2019dualpath}
Y.~Luo, Z.~Chen, and T.~Yoshioka,
\newblock ``Dual-path {RNN}: efficient long sequence modeling for time-domain
  single-channel speech separation,''
\newblock {\em arXiv 1910.06379}, 2019.

\bibitem{isik2016single}
Y.~Isik, J.~Le~Roux, Z.~Chen, S.~Watanabe, and J.~R. Hershey,
\newblock ``Single-channel multi-speaker separation using deep clustering,''
\newblock {\em Interspeech 2016}, pp. 545--549, 2016.

\bibitem{luo2017deep}
Y.~Luo, Z.~Chen, J.~R Hershey, J.~Le~Roux, and N.~Mesgarani,
\newblock ``Deep clustering and conventional networks for music separation:
  Stronger together,''
\newblock in {\em International Conference on Acoustics, Speech and Signal
  Processing (ICASSP)}, 2017, pp. 61--65.

\bibitem{wang2018alternative}
Z.-Q. Wang, J.~Le~Roux, and J.~R. Hershey,
\newblock ``Alternative objective functions for deep clustering,''
\newblock in {\em International Conference on Acoustics, Speech and Signal
  Processing (ICASSP)}, 2018, pp. 686--690.

\bibitem{chen2017deep}
Z.~Chen, Y.~Luo, and N.~Mesgarani,
\newblock ``Deep attractor network for single-microphone speaker separation,''
\newblock in {\em International Conference on Acoustics, Speech and Signal
  Processing (ICASSP)}, 2017, pp. 246--250.

\bibitem{luo2018speaker}
Y.~Luo, Z.~Chen, and N.~Mesgarani,
\newblock ``Speaker-independent speech separation with deep attractor
  network,''
\newblock {\em IEEE/ACM Transactions on Audio, Speech, and Language
  Processing}, vol. 26, no. 4, pp. 787--796, 2018.

\bibitem{kinoshita2018listening}
K.~Kinoshita, L.~Drude, M.~Delcroix, and T.~Nakatani,
\newblock ``Listening to each speaker one by one with recurrent selective
  hearing networks,''
\newblock in {\em International Conference on Acoustics, Speech and Signal
  Processing (ICASSP)}, 2018, pp. 5064--5068.

\bibitem{takahashi2019recursive}
N.~Takahashi, S.~Parthasaarathy, N.~Goswami, and Y.~Mitsufuji,
\newblock ``Recursive speech separation for unknown number of speakers,''
\newblock {\em Interspeech 2019}, pp. 1348--1352, 2019.

\bibitem{buchta2012spherical}
C.~Buchta, M.~Kober, I.~Feinerer, and K.~Hornik,
\newblock ``Spherical k-means clustering,''
\newblock {\em Journal of Statistical Software}, vol. 50, no. 10, pp. 1--22,
  2012.

\bibitem{duda2000pattern}
R.~O. Duda, P.~E. Hart, and D.~G. Stork,
\newblock {\em Pattern classification},
\newblock John Wiley \& Sons, 2000.

\bibitem{hershey2014deep}
J.~R. Hershey, J.~Le Roux, and F.~Weninger,
\newblock ``Deep unfolding: Model-based inspiration of novel deep
  architectures,''
\newblock {\em arXiv preprint arXiv:1409.2574}, 2014.

\bibitem{stoter19}
F.-R. St\"oter, S.~Uhlich, A.~Liutkus, and Y.~Mitsufuji,
\newblock ``Open-unmix - a reference implementation for music source
  separation,''
\newblock {\em Journal of Open Source Software}, 2019.

\bibitem{uhlich2017improving}
S.~Uhlich, M.~Porcu, F.~Giron, M.~Enenkl, T.~Kemp, N.~Takahashi, and
  Y.~Mitsufuji,
\newblock ``Improving music source separation based on deep neural networks
  through data augmentation and network blending,''
\newblock in {\em 2017 IEEE International Conference on Acoustics, Speech and
  Signal Processing (ICASSP)}. IEEE, 2017, pp. 261--265.

\bibitem{kingma2014adam}
D.~P. Kingma and J.~Ba,
\newblock ``Adam: A method for stochastic optimization,''
\newblock {\em arXiv preprint arXiv:1412.6980}, 2014.

\bibitem{SDR_paper}
E.~{Vincent}, R.~{Gribonval}, and C.~{Fevotte},
\newblock ``Performance measurement in blind audio source separation,''
\newblock {\em IEEE Transactions on Audio, Speech, and Language Processing},
  vol. 14, no. 4, pp. 1462--1469, July 2006.

\bibitem{le2019sdr}
J.~Le~Roux, S.~Wisdom, H.~Erdogan, and J.~R. Hershey,
\newblock ``{SDR}--half-baked or well done?,''
\newblock in {\em International Conference on Acoustics, Speech and Signal
  Processing (ICASSP)}, 2019, pp. 626--630.

\end{thebibliography}

\end{document}